# Intanify AI Platform
# Embedded AI for Automated IP Audit and Due Diligence

## Viktor Dörfler[1], Dylan Dryden[2], Viet Lee[2]

[1]University of Strathclyde Business School and Corvinus University of Budapest
[2]Intanify Ltd.

viktor.dorfler@strath.ac.uk, dylan.dryden@intanify.com, viet.lee@intanify.com

**Abstract**

In this paper we introduce a Platform created in order to support SMEs' endeavor to extract value from their intangible assets effectively. To implement the Platform, we developed five knowledge bases using a knowledge-based expert system shell that contain knowledge from intangible asset consultants, patent attorneys and due diligence lawyers. In order to operationalize the knowledge bases, we developed a "Rosetta Stone", an interpreter unit for the knowledge bases outside the shell and embedded in the platform. Building on the initial knowledge bases we have created a system of red flags, risk scoring, and valuation with the involvement of the same experts; these additional systems work upon the initial knowledge bases and therefore they can be regarded as meta-knowledge-representations that take the form of second-order knowledge graphs. All this clever technology is dressed up in an easy-to-handle graphical user interface that we will showcase at the conference. The initial platform was finished mid-2024; therefore, it qualifies as an "emerging application of AI" and "deployable AI", while development continues. The two firms that provided experts for developing the knowledge bases obtained a white-label version of the product (i.e. it runs under their own brand "powered by Intanify"), and there are two completed cases.

# Introduction

In this paper we introduce a deployable AI, an AI-enabled platform, the initial version of which has been recently finished, the first use cases are ongoing (two publicized so far) while the development continues. This means that we have already done the research and development, completed a new AI-enabled platform for intangible asset audit and due diligence (henceforth Platform), but it is not yet sufficiently deployed to submit a case study on deployed application. We discuss our accomplished as well as ongoing efforts to apply AI tools and techniques in novel ways to help companies manage their intangible assets.

As we make our argument for this emerging application, we contribute to the scholarly literature on intangible assets, several aspects of AI, including knowledge engineering, hybrid AI, and human-AI interaction. We also provide insights and a tool for practitioners for managing intangible assets.

The rest of the paper is structured as follows. First, we depict the knowledge background in order to provide the readers with the necessary conceptual framework. This includes arguing for the significance of the intangible asset problem we address, and the relevant considerations of AI technology. The next three sections outline the innovations made along the way, namely process innovations in knowledge engineering, product innovations in the Platform, and content innovations in the way we address intangible assets using AI. We conclude our paper with a Final Commentary, indicating that we are at the beginning of the deployment process; here we note ideas for evaluating the Platform's technical quality and success as well as our future plans for developing the Platform further.

# Background Knowledge

In this section we introduce the conceptual framework needed to understand the nature and significance of the Platform we showcase in this study. We address three areas: (1) Intangible assets (IA) are introduced as the topic of the Platform and their significance today is discussed. (2) AI is introduced in terms of the choice(s) of AI solution(s) adopted for the Platform. (3) Knowledge representations are explored, as part of our innovation relates to these and to the associated knowledge engineering process.

### Intangible Assets

Intangible assets were steadily gaining in significance for the last century or so, by the late 1980s they brought into question established accounting principles (Johnson and Kaplan 1987), and today they are perhaps the most significant topic that companies need to address if they are to be successful (Haskel and Westlake 2018). We start with a simple definition of IA derived from the notion of assets (Lev 2000: 5):

> Assets are claims to future benefits, such as the rents generated by commercial property, interest payments derived from a bond, and cash flows from a production facility. An intangible asset is a claim to future benefits that does not have a physical or financial (a stock or a bond) embodiment. A patent, a brand, and a unique organizational structure (for example, an Internet-based supply chain) that generate cost savings are intangible assets.

Aside the potential for the future value and the lack of physical and financial embodiment, an IA also needs to be identifiable (otherwise we could not even know it existed) and it needs to be within our power to exploit the potential future value (AON and Ponemon Institute 2019: 2).

What makes IA so significant? First, IAs represent huge value, and this value is increasing at a great pace. In the US and Europe alone $3T is invested annually in IA (based on Corrado et al. 2016; see more information at World Intellectual Property Organization 2024). In the market value of S&P 500 companies (the Standard and Poor's 500 stock index tracks the performance of 500 largest companies listed on US stock exchanges) the share of IA increased from 17% in 1975 to 90% in 2020 (see Figure 1). At the beginning of this period the largest of the largest included IBM, GE, P&G, 3M, and Exxon Mobile; by the end of this period all the largest ones were information merchants, such as Google (Alphabet), Apple, Microsoft, Facebook (Meta), Amazon (AON and Ponemon Institute 2019).

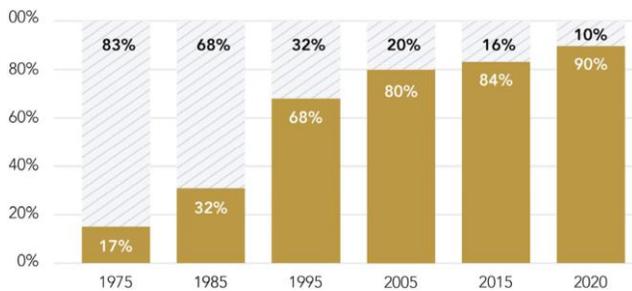

Figure 1: The increase of IA in market capitalization of S&P 500 companies over time (Source: Ocean Tomo 2020)

Second, IA are extremely underutilized. The total value of IA is estimated at $60T worldwide (AON and Ponemon Institute 2022). Typically around 10-15% of the enterprise value is tangible, another 10-15% is covered by IA in the balance sheet (i.e. identified and valued), further 10-15% is identified as asset with unknown value (typically goodwill), and unidentified IA often form about 60% (Hagen 2019). This leads to an underutilization estimated at $1T annually (Forrester Research); most companies cannot even estimate the missed profits or benefits. Due to the exponential trajectory, the situation is continuously getting worse, which makes it paramount for companies to figure out the ways to manage their IAs, which gives significance to our endeavor.

Typically, companies only seek IA consultants and due diligence lawyers' help in the case of mergers and acquisitions. What is the reason for such extreme disregard of potentially huge value for companies? On the one hand, the process is very expensive in terms of both billable hours and invested own labor. Furthermore, the work that needs to be done is not particularly interesting, it involves checklists and spreadsheets, that many employees are tired of, and even the needed interviews are boring, as the process is repetitive (these assertions are underpinned anecdotal evidence of testimonies of people who worked in relevant job roles). On the other hand, many businesses are unaware of what they are missing out.

**Artificial intelligence**

In this paper we do not aim to provide a comprehensive overview of the fast-changing AI literature, we only want to pin down a few concepts that are necessary for our argument. To start with, we use the term AI in a loose and traditional sense, indicating machines that can accomplish tasks humans would do through thinking (see e.g. Dörfler 2020; Guerrero 2023).

With this generic definition in mind and focusing on the underlying technology, we distinguish between three main types of AI (Dörfler 2023a): symbolic reasoning systems, symbolic expert systems, and artificial neural networks.

**Symbolic Reasoning Systems (SRS)**

SRS are systems in which the steps of the problem-solving process, i.e. the steps of reasoning are hard-coded in the system. The steps are typically acquired using "think aloud protocols", meaning that the problem solvers are asked to say aloud every step of thinking they use as they are solving the problem (Dörfler 2022: 11).

Typical examples of SRS include the first known AI solution, the *Logic Theorist* (Newell, Shaw and Simon 1958; Newell, Shaw and Simon 1963), and the famous *Deep Blue* (Finley 2012; Bory 2019) that defeated Garry Kasparov. Today these systems are rare as they are difficult to modify and therefore, they are not flexible. However, they are also extremely robust, therefore we still find them in safety-critical applications, e.g. in airplanes (not the autopilots but e.g. life support) and medical devices (e.g. monitoring systems for comatose patients and medical wearables).

**Symbolic Expert Systems (SES)**

The core of SES a knowledge representation created to reason about the real world (Feigenbaum 1992). The knowledge representation is built by acquiring knowledge from experts in the field through the process of knowledge acquisition conducted by knowledge engineers. The knowledge representation can be programmed directly or built in a software called shell, which includes a reasoning engine as well as an interface for building the knowledge base. When using a shell, the knowledge representation takes the form of a knowledge base (a cleverly constructed database that contains the knowledge representation).

It was Edward Feigenbaum, the student of Herbert Simon and Allen Newell, who figured that the type of AI he envisioned, the later SES, needed a knowledge representation, i.e. an internal model of the external reality, in order to reason about that reality (Feigenbaum 1977). In these early days, Feigenbaum and his colleagues chose "if… then" rules as the form of knowledge representation. Originally these were simple rule lists, but extremely long rule lists destabilized the system quickly on the computers of the late 1960s, therefore Bruce Buchanan, Feigenbaum's long-term collaborator, reprogrammed these into hierarchical structures called "production rules" (Feigenbaum 1992: 11). Essentially, this is still the form of knowledge representation used in SES, the only difference is that sometimes the hierarchy is also presented visually, in the form of a graph, often referred to as rule-based graph (RBG).

Knowledge representations in this sense are unique to SES of all AI types, i.e. these are generic representations of reality from a specific perspective of the expert. This also means that for a different situation (e.g. different decision) but within the same knowledge domain, some of the knowledge base may be reusable.

The generic term for the visual form of knowledge representation is *knowledge graph*. Although there is no complete agreement about the definition of knowledge graphs, it typically indicates a graph of data that represent knowledge in a specific domain, in which the nodes stand for entities that are connected with potentially different relationships, meaning that the nodes stand for things not just data points, and these things can be described using data (Hogan et al. 2021). The first knowledge graphs were those introduced by Google (Singhal 2012), and then followed by the other large companies trading in information, as they were found to be helpful in finding useful information. The term is somewhat generic and some specific graph types, such as rule-based graphs, ID3 graphs, a form of machine learning (ML) in symbolic expert systems (Quinlan 1979, 1986), Bayesian belief networks (conditional probability networks), and many traditional algorithm visualizations can be considered as special cases of knowledge graphs.

What is important for this paper is that production rules can be visualized in the form of knowledge graphs, that multiple knowledge representations can be created using the same nodes with multiple webs connecting the nodes. Furthermore, our future plans include using ML in SES for learning from the experience of using the Platform and thus fine-tuning the knowledge representation based on user experience; this will require the use of a modified ID3, C5 or similar algorithm which we will need to handle together with the RBGs. To this purpose using a single visual representation that is flexible enough to incorporate the various types of knowledge representations is important.

**Artificial neural networks (ANN)**
ANN are without contest the most popular form of AI today; far too often assumed to be the only type. Deep neural networks (DNN) belong under this category, as they are simply ANN with more than one hidden layer (LeCun, Bengio and Hinton 2015).

The process running on ANN is called Machine Learning (ML), even though this is not the only type of machine learning, the name is used almost exclusively. ML that is running on a DNN architecture is called deep learning (DL). ML/DL typically use a very large number of learning examples, in a variant of reinforcement learning, distinguishing between the outcomes of the examples (e.g. good or bad outcome). ANN/DNN address problems by replicating the statistical frequency of responses in specific situations. Notable implementations include AlphaGo, the DeepMind implementation that defeated Lee Sedol (Bory 2019), and AlphaFold, that forecasts a protein structure based on the sequence of amino acids it is made of (Heaven 2020).

In DL the response to the stimulus is produced by the weights assigned to the connections between the artificial neurons, so they do not qualify as knowledge representations in any meaningful sense, although it could be argued that a particular type of distribution of weights in some sense corresponds to some aspect of the situation in the world (Heaven 2019). It is also possible to induce a knowledge representation from an ANN/DNN in the previously noted sense using an ANN-SES hybrid AI.

There are other, lesser known, types of AI that we ignore as they are irrelevant for the current argument, such as genetic algorithms (Holland 1975) or, more generically, evolutionary computing (Fogel, Owens and Walsh 1966), as well as fuzzy logic (Zadeh 1965). These are very useful approaches, sometimes used in combination with the above three, but they are seldom used on their own as AI solutions, the first two belong under optimization techniques while the latter is predominantly used in control engineering, although there are further promising applications (Baranyi 2004; Baranyi, Yam and Várlaki 2017).

**Generative AI (GAI)**
We also did not describe generative AI (GAI), in the previous list of types, because it is not a different technology, it is an approach how technology is used – i.e. it is a different dimension. Along this dimension, AI can be classified as discriminative or generative. Discriminative AI classifies items as belonging or not belonging together, e.g. an image recognition that was trained to recognize cats will group all cats under the label *CAT* and all other images, such as dogs, pianos, bicycles, under the label *NOT CAT*. In contrast, generative AI gets input patterns and then generates an output pattern that is, in some sense, similar to the input ones – but it is not one of the input patterns.

There were early examples of GAI, long before the large language models (LLM). In the 1980s by the composer David Cope created an AI "which could produce very convincing pastiches of a wide range of famous composers" (Boden 2009: 244-245). The important consequence of purpose of GAI is that it is not making a mistake when it "hallucinates", it does exactly what it was programmed to do: it generates a pattern, based on the input patterns, that is similar to the input ones but that did not exist before.

The most recent generation of GAI started with the development of LLM, more precisely, it started with the development of the transformer architecture (Vaswani et al. 2017). With some oversimplification it could be said that having been trained on a large amount of text in the initial stage (pre-training) an LLM gets logic of producing "the next word" in a text. The final stage of training happens in the specific context of the desired application, so "the next word" should be sensible in this context. Of course, as it generates "the next word" it may produce a text that does not make sense, and many implementations today focus on increasing the reliability of GAI in this respect – as it will be noted below, the Platform also deals with this specific problem. Although originally developed in working with text, GAI today is multimodal, we can put in text to prompt generating pictures or the other way around, however for this paper the text aspect of GAI is sufficient.

Importantly for the initial development of the Platform, SES has been chosen as the particular type of AI to be used. This was a non-trivial choice, as DNN is the go-to form of AI today. However, SES was a good choice as we wanted a system that can perform at a high level of mastery (cf Dörfler, Bas and Sinclair forthcoming). SES and DNN require a different setup, that in our case made all the difference. DNN needs a large number of learning examples within a well-defined scope with a reasonable number of variables that are explicit. In contrast, SES does not need a large number of learning examples but it does need input from experts, but we got that covered, early in the process we secured the participation of IA consultants from Mathys & Squire and patent attorneys and due diligence lawyers from Dentons.

## Research Team, Process, and Methods

A remarkably small team produced the AI-enabled Platform showcased in this paper. The team comprised 5 members, including the founders of the company, who are currently the CEO and the CTO; the CEO took the role of the resident decision maker, while the CTO was developing the Rosetta Stone, and implemented all the technical solutions. We had an in-house IA expert with versatile experience in the area of IA. Finally, there were two knowledge engineers, one of them an academic with 25 years of experience in knowledge engineering and a novice, formerly the student of the previous one – not a novice by the end of the project.

In the process research and development were so intertwined, mutually informing each other, that it is hopeless trying to separate them. Our R&D process was cyclic and iterative, featuring "research indirection" (Dörfler, Stierand and Chia 2018), i.e. we learned new things all the time that affected how we proceed, so most project meetings had an aspect of re-planning, we operated in a highly responsive adaptive mode. Remarkably, deliverables were on time.

Methodologically what we did combines the characteristics of participant ethnography (van Maanen 2011) and action research (Eden and Ackermann 2018) design in a critical interpretivist philosophical framing (Dörfler 2023b). The methods of data collection and analysis was SES.

## Knowledge Engineering

The knowledge engineering process is about eliciting experts' knowledge and "recording" it in the software, the SES shell. It is a facilitator-driven computer-supported process (Eden 1992; Dörfler 2021). In the 25-year experience of the knowledge engineer, building a knowledge base typically took 8 working days over a 2-month period, i.e. one day per week. The process usually involved from one to a handful of experts, whose knowledge was elicited, and one or two knowledge engineers performing the elicitation. The preferred setup was two knowledge engineers, as one could then focus on the expert(s) and the other on the computer. Typically, SES were used as decision support tools, meaning that there was the client company that needed decision support, they provided the experts and hired the knowledge engineers. Knowledge engineers in this setup are facilitators, a special kind of consultants with unique skillset (Velencei 2017) and incentives/disincentives (Sims, Eden and Jones 1981).

The outcome of the process is a knowledge base, a representation of the experts' knowledge embedded in the SES software called a shell. The knowledge base is an important object in itself, and should be regarded as a "boundary object", as it is the product of negotiation between the experts and thus at the intersection of their individual perspectives. The knowledge base is also a "transitional object", as it keeps changing throughout the knowledge engineering process (de Geus 1988). The knowledge base is only considered ready when the expert(s) agree that it does correspond to their thinking of the subject. Due to the nature of the knowledge engineering process and of the knowledge base as a boundary and transitional object, the created SES by default satisfies the requirements of explainable AI (Davenport and O'Dell 2019). The nature of the knowledge engineering process also makes the ethical considerations clear; all responsibility is with the experts and the knowledge engineers, AI in this case is simply an amplifier of the ethical considerations (Dörfler and Cuthbert 2024). Importantly, a great deal of precaution is taken in the Platform for the users to understand what the provided AI is capable of, how and for what it can be used.

At the beginning of this project, we established that our situation is substantially different from what is typical. The knowledge engineers worked for the company undertaking the project while the company paid hourly rates to the experts, and four people were assigned to conduct the process of knowledge engineering. This unusual setup resulted in an opportunity to take the knowledge engineering process to a whole new level, resulting in several innovations. The step-change improvement of the knowledge engineering process was one of the intended outcomes of this project, specifically for the scholarly participant. Below we outline these improvements.

There were historical reasons for knowledge engineering to take 8 weeks at one day/week, which is beyond the scope of this paper. At the time, there was no reason to make the process shorter; now, however, we intended to shorten this time significantly, partly as it is typically not affordable today, and partly as the experts' hourly rate was high. From the outset we designed a process that was four days long. To achieve this, a briefing material was designed and sent to the experts before engaging in the knowledge engineering process, so that that they can attend the session prepared. For the same reason, a new example was designed which very quickly explains how the SES works – this is something that experts often struggle with. For the same reason, the two knowledge engineers were there fully engaged all along the process, one focusing on the thinking process of the experts and one focusing on having everything entered into the computer real time, while the knowledge base, as a boundary object, was constantly on the big screen that everyone could see. The in-house expert focused on the content details while the resident decision maker made sure that the knowledge base does not deviate from the scope. These two roles can normally be covered by the knowledge engineers but not in the case when experts are particularly knowledgeable, as then capturing the elusive and nonlinear thought processes takes all the capacity of an experienced knowledge engineer.

Validating a knowledge base is a tedious and time-consuming process, which typically takes place after the completion of the knowledge base. Our particular setup, having an in-house expert, made a huge difference in this. The in-house expert had a relatively broad knowledge of the area, covering the specialty of each expert, while the experts have significantly deeper knowledge. However, this means that the in-house expert was able to quickly notice inconsistencies or anything that would conflict with the generally accepted body of knowledge in the domain. Furthermore, our experts were "grandmasters", i.e. people at the highest level of mastery (Dörfler, Bas and Sinclair forthcoming). Grandmaster knowledge is particularly difficult to elicit, as much of it is tacit and not easily made explicit. It would be fair to say that without the in-house expert much of the grandmaster knowledge could not be unpacked. The in-house expert was able to enter into conversations with the grandmasters teasing out the "essence", i.e. how to change the big picture of the forming knowledge base the way it is needed (Dörfler and Eden 2019: 540-543).

Experts, particularly grandmasters, are not repositories of rules but storytellers of their former experience. Traditionally, rules are often more or less finalized before entering any cases, to test those rules. In the current project it has proven exceptionally useful to switch to adding cases of experience several times, this provided opportunity for questions like "what makes case X so different from case Y, that warrants a significantly better outcome?" Interesting cases often also remind experts of less obvious attributes (variables), and it is also very useful if this does not happen in the end, when it can lead to a major change of the entire knowledge base. The ongoing validation and the iterative use of rules and cases helped reduce the knowledge engineering process to three days.

Next, one of the knowledge engineers and the in-house expert prepared rapid prototype knowledge bases using the established knowledge in the literature and these have been sent to the experts beforehand. Importantly, the experts, who were grandmasters, often significantly changed these draft knowledge bases, however it was something to work with, and the knowledge engineering process was reduced to two days. On one occasion, with an expert who was particularly knowledgeable as well as exceptionally quick to pick up the specific way of thinking needed for knowledge engineering, we actually managed to create a full knowledge base in a single day. With this, we believe we have shortened the process as much as possible, in fact the one-day version will rarely be possible with the current technology. However, the experience gave us excellent ideas on how to develop a technology that can make the one-day knowledge engineering a standard; this is, however, matter of future software development in the area of SES.

## AI Embedded in a Platform

Software in which SES are built, the so-called shells, are extremely useful tool for knowledge engineers, but they are not fit for end users. From the outset, we wanted an AI-enabled platform that meets commercial expectations, so that the users do not need to use anything else. For that we had to move the knowledge base outside the shell and into the Platform. There are two ways to do this, either the shell can export the knowledge base in a convenient format, or a new component can be built that "translates" the knowledge base into computer code that the Platform can run for users to interact with. We opted for the second variant, and built what we refer to as the "Rosetta Stone" of our Platform; it reads the standard knowledge base, parses it and produces Typescript code in the Platform schema. This means that instead of navigating the spreadsheet-like UI of the shell, with many unnecessary (for the user) functions, the users access a clean direct interface, like the one of Figure 2.

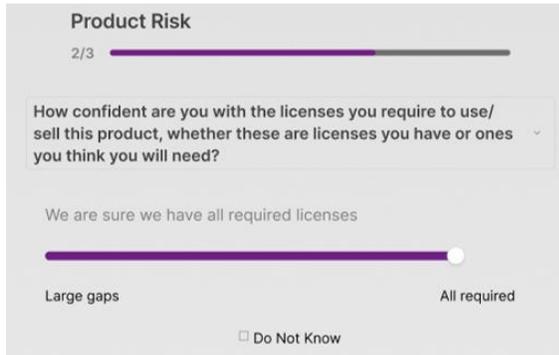

Figure 2: using the knowledge base in the platform

We skip describing the technical difficulties that came with the initial programming, the innovation here was really the idea of creating the Rosetta Stone. However, as we had complete control of the Rosetta Stone, we could also do minor amendments of the knowledge bases in here, and also develop external connectors to it. Another very interesting innovation was producing *layers of reasoning* grounded in the same knowledge representation. For instance, we were able to add risk scores, as not all decisions were equal, and red flags, as combinations of states come together from different parts of the knowledge base signifying specific uses, that can be threats or opportunities.

Once the initial knowledge bases were transferred into the Platform, we have finetuned the language (i.e. the displayed wording). The grandmasters' knowledge is often metaphoric and always very contextual; therefore, some initial rewording was needed. This was done by the in-house expert and the knowledge engineer; the in-house expert made sure that the wording is professionally completely correct and he explained it to the knowledge engineer and only when she fully understood it the knowledge base was considered to be "second draft" in terms of wording.

In the next instance, the knowledge base, now embedded in the platform, was given to beta testers, who were representatives of the future users of the platform. Very quickly it was realized that additional explanations and longer wordings may be needed, although this was not even across all the beta-tester experience. In order to provide as much information as needed but only to those who needed it, the Rosetta Stone was equipped with a hierarchical explanatory content, which can be thought of as a contextual drill-down help feature. Once the wording was working well across the full tester population, we have looked into speeding up the user experience too. For instance, a library of sliders was connected to the Rosetta Stone to graphically display users' choices of input.

Similarly, GUI elements are generated from the information on the output side, producing a user-friendly dashboard, e.g., Figure 3 displays IA by categories, supporting IA discovery.

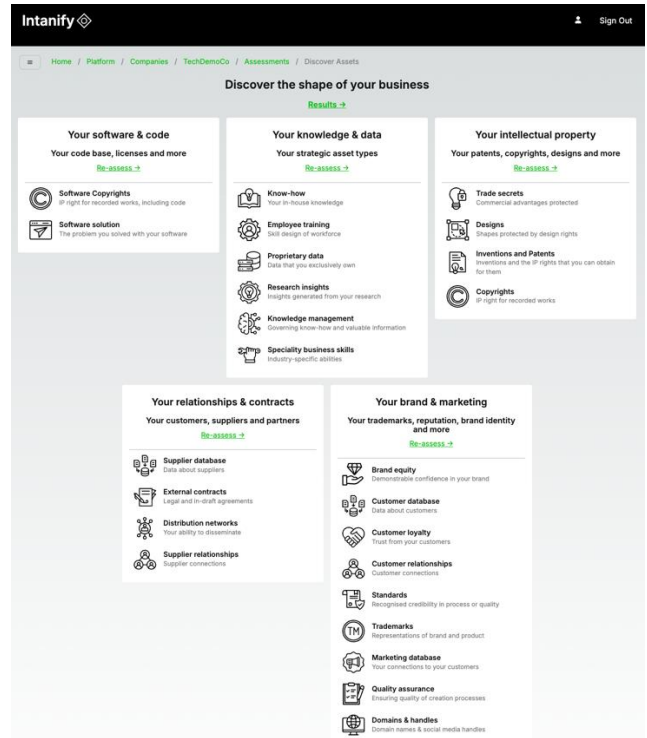

Figure 3: Shape of business dashboard from IA.

The most interesting technical opportunity building on the Rosetta Stone, however, is a development of a SES-GAI hybrid AI. There are three uses of such hybrid AI that we currently have in test deployment:

- GAI can partly populate the input to the Platform with data extracted from documents that are provided by the company or publicly available.
- GAI can generate reports on IA, within a scope tightly controlled by SES and from the data produced in the Platform.
- Finally, SES can be used to provide prompts to GAI, which is can help answer user queries.

Another area of future development is to offer flexible Platform plug-ins to the users, currently it can be used as a tool, however, it could also feed information produced by the Platform into various decision tools, like an "AI assistant", or could scrutinize financial decisions as an "AI peer collaborator" (cf Göndöcs and Dörfler 2024: 6-7).

## AI for Managing IA

The content innovation aspect of the project is perhaps more important than the process and technical innovations presented in the previous two sections and what we show here also includes additional process and technical innovations – but the focus is on the content. There are two such content innovations implemented, one concerning red flags and another one concerning valuation and risk quantification.

In the knowledge engineering process, we have built 5 knowledge bases, comprising nearly 200 input attributes (variables), 75 on IP audit, 12 on valuation, and 100+ on due diligence, resulting in approximately 12,000 rules. And this number means hierarchically organized production rules, which correspond to about 100,000 rules if they were done as single-level rule lists. Once the initial knowledge bases were ready, we discussed specific cases with the experts and started asking them about what were the red flags that indicated e.g. that Client $X$ had a problem $Y$. We quickly realized that the experts were pointing at various parts of the knowledge bases, but not in the way these were structured. Their intuitive alarm went off for particular constellations of various parts of the knowledge bases – we have mapped these patterns that operated in a sense 'on the top' of the original knowledge bases, providing us with the opportunity to construct layers of reasoning. We cannot share these as they are the IA of Intanify.

Next using the combination of red flags and the knowledge bases, we discussed with the experts the estimation process of quantified risk; this is another second-layer reasoning but as it operates with quantities, this was done in an Excel table, before replicating it in the Platform.

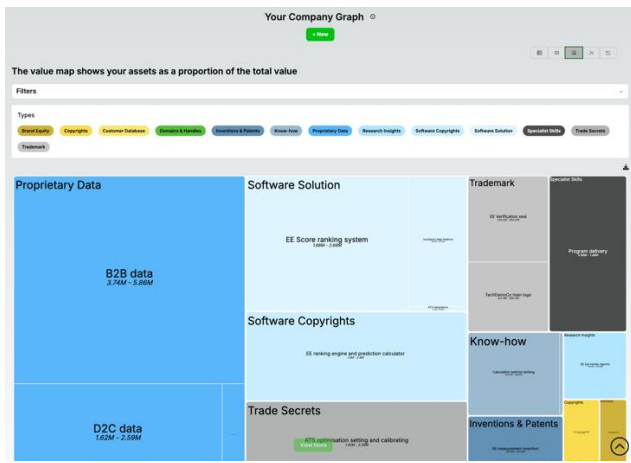

Figure 4: Valuation of IA.

---

[1] Intanify case published 19th November 2024: Uncovering Potential – How Orderly Positioned for Funding Success, available online at: https://insights.intanify.com/orderly-intanify-0

Using a similar logic, we have developed a valuation (Figure 4) module of the of the dashboard; the box sizes here are proportionate with the relative IA values.

Overall, the Platform automates expertise acquired from top experts in the field and applies it to the data that the users enter about their companies. It is a powerhouse of embedded knowledge, including 12,000 qualitative rules, full risk and valuation procedures and flagging red flags that send the users to see IA consultants or lawyers if needed.

These innovations together led to a unique user value. Intanify publicizes a case study of Orderly, a company with 51-200 (undisclosed) employees and a 75% growth in 2023, currently preparing for its Series-A in 2025 funding round[1]. They strengthened their position by developing a robust IP portfolio and mitigating risk through the identification of 73 key assets which underpin the revenue, growth and moat of the business.

"Orderly now better understands and can demonstrate how its IP assets support its success, while also adjusting its strategy to mitigate risks and maximise potential. Ultimately, this is a win-win, it's a win for the business and it's a win for future investors." (Dr. Justin Hill, Partner and Co-Chair of Intellectual Property at Dentons)

## Final Commentary

In this paper we have introduced an AI-enabled portal that can automate basic IA audit, including identification of IA in 25 asset types, valuation, risk assessment, and a variety of insights, including red flags. It helps users understand the value of their IA and the associated risks without extremely expensive auditing or legal advice. The small amount of user feedback we have is overwhelmingly positive.

Furthermore, the part of the process we have automated is what IA consultants, IP attorneys, and due diligence lawyers are also not particularly keen on as it is all about the trivial cases that are neither interesting nor profitable. This is probably the reason that the two companies that provided us with experts, Mathys & Squire and Dentons both acquired a white-label version of the Platform to integrate it into their business processes (i.e. it runs under their own brand "powered by Intanify"), and used it with great success:

"It's not often you can have a call where the other party immediately understands what your business priorities are and what drives you. Dentons knew instantly where the challenges were and then recommended a strategy within minutes." (Peter Evans, CEO at Orderly)

## Acknowledgments

We are grateful for the Innovate UK grant (10052286) that funded the research project and made the industry-academy collaboration possible.